\documentclass[aps,prb,twocolumn,showpacs]{revtex4}
\usepackage{graphicx,amsmath,amssymb,bm}
\begin{document}

\title{Infrared catastrophe and tunneling into strongly correlated electron systems: Beyond the x-ray edge limit}
\author{Kelly R. Patton and Michael R. Geller}
\affiliation{Department of Physics and Astronomy, University of Georgia, Athens, Georgia 30602-2451}

\date{September 23, 2005}

\begin{abstract}
We develop a nonperturbative method to calculate the electron propagator in low-dimensional and strongly correlated electron systems. The method builds on our earlier work using a Hubbard-Stratonovich transformation to map the tunneling problem to the x-ray edge problem, which accounts for the infrared catastrophe caused by the sudden introduction of a new electron into a conductor during a tunneling event. Here we use a cumulant expansion to include fluctuations about this x-ray edge limit. We find that the dominant effect of electron-electron interaction at low energies is to correct the noninteracting Green's function by a factor $e^{-S/\hbar}$, where $S$ can be interpreted as the Euclidean action for a density field describing the time-dependent charge distribution of the newly added electron. Initially localized, this charge distribution spreads in time as the electron is accommodated by the host conductor, and during this relaxation process action is accumulated according to classical electrostatics with a screened 
interaction. The theory applies to lattice or continuum models of any dimensionality, with or without translational invariance. In one dimension the method correctly predicts a power-law density of states (DOS) for electrons with short-range interaction and no disorder, and when applied to the solvable Tomonaga-Luttinger model, the exact DOS is obtained. 
\end{abstract}


\maketitle
\section{Introduction}
In the conventional many-body theory treatment of tunneling, the low-temperature tunneling current between an ordinary metal and a strongly correlated electron system is controlled by the single-particle density of states (DOS) of the correlated system. Tunneling experiments are therefore often used as a probe of the DOS. In a wide variety of low dimensional and strongly correlated electron systems, including all 1D metals, the 2D diffusive metal, the 2D Hall fluid, and the edge of the sharply confined Hall fluid, the DOS exhibits anomalies such as cusps, algebraic suppressions, and pseudogaps at the Fermi energy. We have recently proposed that the origin of these anomalies is the infrared catastrophe caused by the response of the host electron gas to the sudden introduction of a new particle that occurs during a tunneling event.\cite{Patton&GellerPRB05} 

The infrared catastrophe is a singular screening response of a degenerate Fermi gas to a localized potential applied abruptly in time, which is known to be responsible for the anomalous x-ray optical and photoemission edge spectra of metals,\cite{MahanPR67,Nozieres&DeDominicisPR69} the Anderson orthogonality catastrophe,\cite{AndersonPRL67,HamannPRL71} and the Kondo effect.\cite{AndersonPRL69,YuvalPRB70} Imagine an electron in a tunneling process being replaced by a negatively charged, distinguishable particle with mass $M$. In the $M \! \rightarrow \! \infty$ limit, the time-dependent potential produced by the particle being added to the origin at time $\tau = 0$  and removed at a later time $\tau_0$ is\cite{stepfunctionnote}
\begin{equation}
\phi_{\rm xr}({\bf r},\tau) = U({\bf r}) \Theta(\tau) \Theta(\tau_{0}-\tau),
\label{phixr}
\end{equation}
where $U$ is the two-particle interaction. $\phi_{\rm xr}$ is identical, up to a sign, to the abruptly turned-on hole potential of the x-ray edge problem, so an infrared catastrophe would be expected. Tunneling of a real, finite-mass electron is different because it recoils, softening the potential produced. The real electron is also an indistinguishable fermion, unable to tunnel into the occupied states below the Fermi energy.

In Ref.~\onlinecite{Patton&GellerPRB05} we introduced an exact functional-integral representation for the interacting propagator, and developed a nonperturbative technique for calculating it by identifying a ``dangerous" scalar field configuration of the form (\ref{phixr}), and then treating this special field configuration by using methods developed for the x-ray edge problem. All other field configurations were ignored, thereby reducing the tunneling problem to an x-ray edge problem. Nonetheless, qualitatively correct results were obtained for the 1D electron gas and the 2D Hall fluid using this approach. In this paper we attempt to go beyond this so-called x-ray edge limit by including fluctuations about $\phi_{\rm xr}$. We find that by including fluctuations through the use of a simple functional cumulant expansion, a qualitatively correct DOS is obtained for electrons with short-range interaction and no disorder in one, two, and three dimensions. We also show that and when applied to the solvable Tomonaga-Luttinger model, the low-energy fixed-point Hamiltonian for most 1D metals, the exact DOS is obtained. A preliminary account of this work has appeared elsewhere.\cite{Geller02}

\section{general formalism and cumulant expansion method}\label{general formalism section} 

We assume a grand-canonical Hamiltonian $H =H_0 + V$, where
\begin{equation}
H_0 \equiv \sum_\sigma \int d^{\scriptscriptstyle D}r \ \psi^\dagger_\sigma({\bf r}) \, \bigg[ {\Pi^2 \over 2m} + v({\bf r}) - \mu \bigg] 
\psi_\sigma ({\bf r})
\end{equation}
is the Hartree mean-field Hamiltonian, and
\begin{equation} 
V \equiv {\textstyle{1 \over 2}} \int d^{\scriptscriptstyle D}r \ d^{\scriptscriptstyle D} r' \, \delta n({\bf r}) \, U({\bf r}-{\bf r}') \, \delta n({\bf r}').
\label{interaction hamiltonian}
\end{equation}
Here ${\bf \Pi} \equiv {\bf p} + {\textstyle{e \over c}} {\bf A}$, and $v({\bf r})$ is any single-particle potential, which may include a periodic lattice potential or disorder or both, and which also includes the Hartree interaction with the self-consistent density $n_0({\bf r})$. The interaction term is written in terms of the density fluctuation
\begin{equation}
\delta n({\bf r}) \equiv \sum_\sigma \psi^\dagger_\sigma({\bf r}) 
\psi_\sigma({\bf r}) \ - \ n_0({\bf r}).
\end{equation}
Following Ref.~\onlinecite{Patton&GellerPRB05}, we use a Hubbard-Stratonovich transformation to write the exact Euclidean propagator 
\begin{equation}
G({\bf r}_{\rm f}\sigma_{\rm f},{\bf r}_{\rm i} \sigma_{\rm i},\tau_0) \equiv - \big\langle T \psi_{\sigma_{\rm f}}({\bf r}_{\rm f} 
, \tau_0) {\bar \psi}_{\sigma_{\rm i}}({\bf r}_{\rm i} ,0) \big\rangle_{\! H},
\label{G definition}
\end{equation}
in the form
\begin{equation}
G({\bf r}_{\rm f} \sigma_{\rm f}, {\bf r}_{\rm i} \sigma_{\rm i},\tau_0) = {\cal N} \, {\int \! D\phi \, e^{-{1 \over 2} \!  \int \! \phi {U}^{-1} \! \phi} 
g({\bf r}_{\rm f} \sigma_{\rm f}, {\bf r}_{\rm i} \sigma_{\rm i}, \tau_0 | \phi) \over \int \! D\phi \, e^{-{1 \over 2} \! \int \! \phi {U}^{-1} \! \phi}} 
\label{exact form}
\end{equation}
where
\begin{equation}
g(\phi) \equiv  - \big\langle T \psi_{\sigma_{\rm f}}({\bf r}_{\rm f},\tau_0)
{\bar \psi}_{\sigma_{\rm i}}({\bf r}_{\rm i},0) \, e^{i \int \! \phi({\bf r},\tau) \, \delta n({\bf r},\tau) } \big\rangle_0 
\label{correlation function definition}
\end{equation}
is a noninteracting correlation function in the presence of a purely imaginary scalar potential $i\phi({\bf r},\tau)$, and ${\cal N} \equiv \langle T 
\exp(-\int_0^\beta d \tau \, V) \rangle_0^{-1}$ is a constant, independent of $\tau_{0}$. Eq.~(\ref{exact form}) is an exact expression for the interacting Green's function. 

The region of function space that contributes to the functional integrals in (\ref{exact form}) is controlled by the width of the Gaussian, which in the small $U$ limit becomes strongly localized around $\phi=0$. By expanding (\ref{correlation function definition}) in powers of $\phi$ and doing the functional integrals term by term, one simply recovers the standard perturbative expansion for $G({\bf r}_{\rm f}\sigma_{\rm f},{\bf r}_{\rm i} \sigma_{\rm i},\tau_0)$ in powers of $U$. Therefore, it will be necessary to go beyond a perturbative expansion for $g({\bf r}_{\rm f} \sigma_{\rm f}, {\bf r}_{\rm i} \sigma_{\rm i},\tau_0 | \phi)$. We evaluate $g({\bf r}_{\rm f} \sigma_{\rm f}, {\bf r}_{\rm i} \sigma_{\rm i},\tau_0 | \phi)$ approximately, using a second-order functional cumulant expansion. Such an expansion amounts to a resummation of the most divergent terms in the perturbation series when $\phi=\phi_{\rm xr}$ and the infrared catastrophe occurs. Indeed, one can view our resulting expression for $g({\bf r}_{\rm f} \sigma_{\rm f}, {\bf r}_{\rm i} \sigma_{\rm i},\tau_0 | \phi)$ as a functional generalization of Mahan's ``perturbative" result for a similar correlation function.\cite{MahanPR67} Furthermore, for field configurations far from $\phi_{\rm xr}$, the cumulant expansion will yield a result that is, by construction, exact through second order in $U$. After carrying this out we obtain 
\begin{eqnarray}
g(\phi) &\approx& G_0({\bf r}_{\rm f} \sigma_{\rm f}, {\bf r}_{\rm i} \sigma_{\rm i}, \tau_0) \nonumber \\
&\times&  e^{ \int C_1({\bf r} \tau) \, \phi({\bf r},\tau) + \int  C_2({\bf r} \tau, {\bf r}' \tau' ) \, \phi({\bf r},\tau) \, \phi({\bf r}', \tau') },
\label{cumulant expansion}
\end{eqnarray}
where
\begin{equation}
C_1({\bf r} \tau) = { g_1({\bf r} \tau) \over 
G_0({\bf r}_{\rm f} \sigma_{\rm f},{\bf r}_{\rm i} \sigma_{\rm i},\tau_0)}
\end{equation}
and
\begin{equation}
C_2({\bf r} \tau, {\bf r}' \tau' ) = { g_2({\bf r} \tau , {\bf r}' 
\tau') \over 
G_0({\bf r}_{\rm f} \sigma_{\rm f},{\bf r}_{\rm i} \sigma_{\rm i},\tau_0)} -
{ g_1({\bf r} \tau) \, g_1({\bf r}' \tau')  \over 2 \, 
[G_0({\bf r}_{\rm f} \sigma_{\rm f},{\bf r}_{\rm i} \sigma_{\rm i},\tau_0)]^2}.
\end{equation}
Here 
\begin{align}
&g_n({\bf r}_1 \tau_1 , \cdots , {\bf r}_n \tau_n) \equiv \nonumber\\ &- \, 
{i^n \over n!} \, \big\langle T \psi_{\sigma_{\rm f}}({\bf r}_{\rm f} ,\tau_0)
{\bar \psi}_{\sigma_{\rm i}}({\bf r}_{\rm i} ,0) \, \delta n({\bf r}_1 \tau_1) \,
\cdots \, \delta n({\bf r}_n \tau_n) \big\rangle_0 
\end{align}
is the coefficient of $\phi^n$ appearing in the perturbative expansion of 
(\ref{correlation function definition}), as in
\begin{align}
g&({\bf r}_{\rm f} \sigma_{\rm f}, {\bf r}_{\rm i} \sigma_{\rm i},\tau_0
| \phi) =  G_0({\bf r}_{\rm f} \sigma_{\rm f}, {\bf r}_{\rm i} \sigma_{\rm i},
\tau_0) \nonumber \\ +&\sum_{n=1}^\infty \int g_n({\bf r}_1 \tau_1, \cdots, 
{\bf r}_n \tau_n) \, \phi({\bf r}_1,\tau_1) \cdots \phi({\bf r}_n,\tau_n).
\label{formal perturbation expansion}
\end{align}
The cumulants $C_1$ and $C_2$ in terms of $G_{0}$ are 
\begin{equation}
C_1({\bf r} \tau) = - i\sum_\sigma
{G_0({\bf r}_{\rm f} \sigma_{\rm f}, {\bf r} \sigma , \tau_0 - \tau) \, 
G_0({\bf r} \sigma , {\bf r}_{\rm i} \sigma_{\rm i}, \tau) \over 
G_0({\bf r}_{\rm f} \sigma_{\rm f}, {\bf r}_{\rm i} \sigma_{\rm i}, \tau_0)}  
\end{equation}
and (suppressing spin for clarity)
\begin{align}
&C_2({\bf r} \tau,{\bf r}' \tau')=\frac{1}{2G_0({\bf r}_{\rm f}, {\bf r}_{\rm i}, \tau_0)^{2}}\nonumber \\ & \times\Big\{[G_0({\bf r}, {\bf r}', \tau-\tau')G_0({\bf r}_{\rm f}, {\bf r}_{\rm i}, \tau_0)\nonumber \\ & -G_0({\bf r}, {\bf r}_{\rm i}, \tau)G_0({\bf r}_{\rm f}, {\bf r}', \tau_0-\tau')][{\bf r}\leftrightarrow {\bf r}',\tau \leftrightarrow\tau']\Big\}
\end{align}

The functional integral in (\ref{exact form}) can now be done exactly, leading to
\begin{equation}
G({\bf r}_{\rm f} \sigma_{\rm f}, {\bf r}_{\rm i} \sigma_{\rm i},\tau_0)
= {\cal A}(\tau_0) \ G_0({\bf r}_{\rm f} \sigma_{\rm f}, {\bf r}_{\rm i} 
\sigma_{\rm i},\tau_0) \ e^{-S(\tau_0)}, 
\label{general result}
\end{equation}
where
\begin{align}
{\cal A} & \equiv {\cal N} \, {\int D\phi \ e^{-{1 \over 2} \int \phi ({ U}^{-1}
- 2 C_2) \phi} \over \int D\phi \ e^{-{1 \over 2} \int \phi { U}^{-1} \phi}}
\nonumber \\ &=   {\cal N} \big[ \det\, (1-2 C_2 { U}) \big]^{-{1 \over 2}} , 
\label{A definition}
\end{align}
and
\begin{equation}
S \equiv {1 \over 2} \int_0^\beta \! \! d\tau \, d\tau' \!  \int \! d^{\scriptscriptstyle D}r \, d^{\scriptscriptstyle D}r' \rho({\bf r},\tau) \, 
U_{\rm eff}({\bf r} \tau, {\bf r}' \tau') \, \rho({\bf r}',\tau'). 
\label{action}
\end{equation}
Here
\begin{align}
&\rho({\bf r},\tau) \equiv -i \, C_1({\bf r} \tau)  \\& = - \sum_\sigma {G_0({\bf r}_{\rm f} \sigma_{\rm f}, {\bf r} \sigma , \tau_0 - \tau) \, 
G_0({\bf r} \sigma , {\bf r}_{\rm i} \sigma_{\rm i}, \tau) \over G_0({\bf r}_{\rm f} \sigma_{\rm f}, {\bf r}_{\rm i} \sigma_{\rm i}, \tau_0)}  ,
\label{tunneling charge definition}
\end{align}
and
\begin{equation}
U_{\rm eff}({\bf r} \tau, {\bf r}' \tau') \equiv  \big[{ U}^{-1}({\bf r}- {\bf r}') \delta(\tau-  \tau') - 2 C_2({\bf r} \tau, {\bf r}' \tau') 
\big]^{-1}
\label{Ueff definition}
\end{equation}
is a screened interaction.

Because spin-orbit coupling has been neglected in $H$, the noninteracting Green's function is diagonal in spin, and
\begin{eqnarray}
\rho({\bf r},\tau) = - 
{G_0({\bf r}_{\rm f} \sigma_{\rm i}, {\bf r} \sigma_{\rm i}, \tau_0 - \tau) \, G_0({\bf r} \sigma_{\rm i} , {\bf r}_{\rm i} \sigma_{\rm i}, \tau) \over 
G_0({\bf r}_{\rm f} \sigma_{\rm i}, {\bf r}_{\rm i} \sigma_{\rm i}, \tau_0)} \ \delta_{\sigma_{\rm i} \sigma_{\rm f}} .
\label{tunneling charge with spin conserved}
\end{eqnarray}
Eq.~(\ref{general result}) is the principal result of this work. 

\section{CHARGE SPREADING INTERPRETATION}

We interpret (\ref{general result}) as follows: $S$ is the Euclidean action\cite{functionalnote} for a time-dependent charge distribution $\rho({\bf r},\tau)$. We shall show that $\rho({\bf r},\tau)$ acts like a charge density associated with an electron being inserted at ${\bf r}_{\rm i}$ at $\tau \! = \! 0$ and removed at ${\bf r}_{\rm f}$ at $\tau_0$. This charge density interacts via an effective interaction $U_{\rm eff}({\bf r} \tau, {\bf r}' \tau')$ that accounts for the modification of the electron-electron interaction by dynamic screening.\cite{photonfootnote} Our result can therefore be regarded as a variant of the intuitive but phenomenological ``charge spreading'' picture of Spivak\cite{spivak} and of Levitov and Shytov.\cite{Levitov97} However, here the dynamics of  $\rho({\bf r},\tau)$ is completely determined by the mean-field Hamiltonian, and has the dynamics of essentially noninteracting electrons.

First consider the integrated charge,
\begin{eqnarray}
Q(\tau) &\equiv& \int d^{\scriptscriptstyle D}r \ \rho({\bf r},\tau) \nonumber \\
&=& -\frac{\int d^{\scriptscriptstyle D}r\ G_{0}({\bf r}_{\rm f},{\bf r},\tau_{0}-\tau)G_{0}({\bf r},{\bf r}_{\rm i},\tau)}{G_{0}({\bf r}_{\rm f},{\bf r}_{\rm i},\tau_{0})}.
\label{integrated charge}
\end{eqnarray}
Using exact eigenfunction expansions for the Green's functions we obtain

\begin{widetext}

\begin{equation}
Q(\tau)= -\frac{\displaystyle\sum_{\alpha}\Phi^{*}_{\alpha}({\bf r}_{\rm f})\Phi_{\alpha}({\bf r}_{\rm i})e^{-(\epsilon_{\alpha}-\mu)\tau_{0}}
\left\{[n_{\rm F}(\epsilon_{\alpha}-\mu)-1]^{2}\Theta(\tau_{0}-\tau)\Theta(\tau)+n_{\rm F}(\epsilon_{\alpha}-\mu)[n_{\rm F}(\epsilon_{\alpha}-\mu)-1]
\left[\Theta(-\tau)+\Theta(\tau-\tau_{0})\right]\right\}}{\displaystyle\sum_{\alpha}\Phi^{*}_{\alpha}({\bf r}_{\rm f})\Phi_{\alpha}({\bf r}_{\rm i})e^{-(\epsilon_{\alpha}-\mu)\tau_{0}}[n_{\rm F}(\epsilon_{\alpha}-\mu)-1]}.
\end{equation}
where $n_{\rm F}$ is the Fermi distribution function and the $\Phi_\alpha$ are the single-particle eigenfunctions of $H_0$. In the zero temperature limit
\begin{equation}
Q(\tau)=-\frac{\displaystyle\sum_{\alpha}\Phi^{*}_{\alpha}({\bf r}_{\rm f})\Phi_{\alpha}({\bf r}_{\rm i})e^{-(\epsilon_{\alpha}-\mu)\tau_{0}}\left\{(N_{\alpha}-1)^{2}\Theta(\tau_{0}-\tau)\Theta(\tau)+N_{\alpha}(N_{\alpha}-1)\left[\Theta(-\tau)+
 \Theta(\tau - \tau_{0})\right]\right\}}{\displaystyle\sum_{\alpha}\Phi^{*}_{\alpha}({\bf r}_{\rm f})\Phi_{\alpha}({\bf r}_{\rm i})e^{-(\epsilon_{\alpha}-\mu)\tau_{0}}(N_{\alpha}-1)},
\label{general sum rule}
\end{equation}
\end{widetext}
where $N_{\alpha}$ is the ground-state occupation number of state $\alpha$, which in the absence of ground state degeneracy takes the value of $0$ or $1$. In this case (\ref{general sum rule}) reduces to
\begin{equation}
Q(\tau) = \Theta(\tau_0 - \tau)\Theta(\tau).
\label{sum rule}
\end{equation} 
When the sum rule (\ref{sum rule}) holds, the net added charge, as described by $\rho({\bf r},\tau)$, is unity (in units of the electron charge) for times between $0$ and $\tau_{0}$, and zero otherwise. This behavior correctly mimics the action of the field operators in (\ref{G definition}).

\begin{figure}
\includegraphics[width=8.0cm]{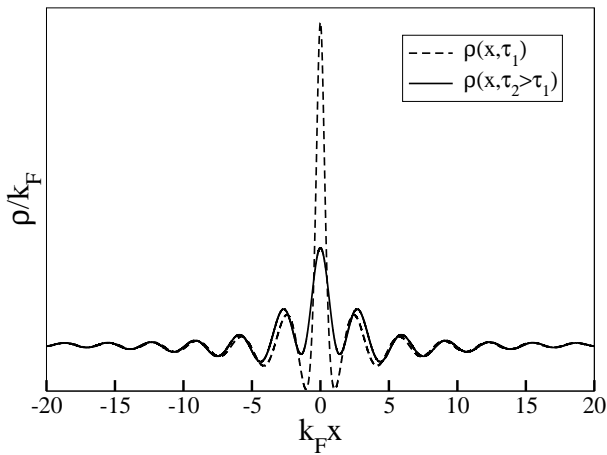}
\caption{\label{rhofig} Charge density $\rho (x,\tau)$ for the 1D electron gas at two times, showing Friedel oscillations and gradual spreading.}
\end{figure}

At short times, $\tau \ll \tau_0$, the charge density is approximately
\begin{equation}
\rho({\bf r},\tau) \sim -G_{0}({\bf r},{\bf r}_{\rm i},\tau)
\end{equation}
which is localized around ${\bf r}={\bf r}_{\rm i}$. As time evolves this distribution relaxes. Then as $\tau$ approaches $\tau_{0}$ the charge density again becomes localized around ${\bf r}={\bf r}_{\rm f},$ 
\begin{equation}
\rho({\bf r},\tau)\sim -G_{0}({\bf r}_{\rm f},{\bf r},\tau_0 -\tau).
\end{equation}
A plot of $\rho({\bf r},\tau)$ for the 1D electron gas is given in Fig.~\ref{rhofig}.
 
The dynamics of $\rho({\bf r},\tau)$ can be shown to be governed by the equation of motion
\begin{equation}
\left[\partial_{\tau}+H_{0}({\bf r})\right]\rho({\bf r},\tau)=-\delta({\bf r}_{\rm f}-{\bf r})\delta({\tau}_{0}-\tau')+\delta({\bf r}-{\bf r}_{\rm i})\delta({\tau}).
\label{equation of motion}
\end{equation}
This can be seen by noting that the noninteracting Green's function satisfies 
\begin{equation}
\left[\partial_{\tau}+H_{0}({\bf r})\right]G_{0}({\bf r},{\bf r'},\tau,\tau')=-\delta({\bf r}-{\bf r}')\delta({\tau}-\tau').
\end{equation}
Then, using the definition (\ref{tunneling charge definition}) one can obtain (\ref{equation of motion}). Again, we stress that $\rho({\bf r},\tau)$ is describing the dynamics of noninteracting electrons, governed by the mean-field Hamiltonian $H_0$.
 
Although $\rho({\bf r},\tau)$ has many properties that make it reasonable to interpret as the charge density associated with the added and subsequently removed electron in the Green's function (\ref{G definition}), one should not take this interpretation too literally.  For instance, the sum rule (\ref{sum rule}) only holds in the zero temperature limit and in the absence of ground state degeneracy. Also, as will be seen below in Sec.~\ref{TL model section}, $\rho({\bf r},\tau)$ may even be complex valued. 
 
\section{APPLICATIONS OF THE CUMULANT METHOD} 

In the following examples we will assume electrons with a short-range interaction $U$, no disorder, and no magnetic field. In Sec.~\ref{fermi liquid section} we show that our method correctly predicts a constant DOS near the Fermi energy in 2D and 3D, and in Sec.~\ref{luttinger liquid section} we obtain a power-law DOS in 1D, in qualitative agreement with Luttinger liquid theory.\cite{TomonagaPTP50,MattisJMP65,DzyaloshinskiiJETP74,HaldaneJPC81,HaldanePRL81,KanePRB92,Fisher97} Finally, in Sec.~\ref{TL model section} we use our method to calculate the DOS for the solvable Tomonaga-Luttinger model, obtaining the exact DOS exponent.

\subsection{2D and 3D electron gas: Recovery of the Fermi liquid phase}\label{fermi liquid section}

The sum rule (\ref{sum rule}) allows one to determine the energy dependence of the DOS in $D$ dimensions, asymptotically in the low-energy limit, as follows: In the absence of disorder, a droplet of charge injected into a degenerate Fermi gas with velocity $v_{\rm F}$ will relax to a size of order $\ell \sim v_{\rm F} \tau$ after a time $\tau$. Approximating $\rho({\bf r},\tau)$ to be of uniform magnitude in a region of size $\ell$ and vanishing elsewhere, the sum rule then requires the magnitude or $\rho$ to vary as $\ell^{-D} \! .$ The interaction energy of such a charge distribution (assuming a short-range interaction) is 
\begin{equation}
E = {U \over 2} \int d^{\scriptscriptstyle D}r \, [\rho({\bf r})]^2 \sim \frac{U}{\ell^D}, 
\end{equation}
which varies with time as $\tau^{-D}$. The action accumulated up to time $\tau_0$ therefore scales as 
\begin{equation}
S \sim \frac{U}{\tau_0^{D-1}} \ \ \ \ {\rm if} \ \ \ \ D\ge2 
\label{FL spreading}
\end{equation}
or
\begin{equation}
S \sim U \ln \tau_0 \ \ \ \ {\rm if} \ \ \ \ D=1. 
\label{LL spreading}
\end{equation}

The cases (\ref{FL spreading}) and (\ref{LL spreading}) are dramatically different: In 2D and 3D the action vanishes a long times, and the propagator (\ref{general result}) is therefore not appreciably affected by interactions. The resulting DOS is energy independent at low energies, and the expected Fermi liquid behavior is recovered. In 1D, however, the action diverges logarithmically, leading to an algebraic DOS.  

\subsection{1D electron gas: Recovery of the Luttinger liquid phase}\label{luttinger liquid section}

The scaling argument of the previous section showed that the DOS in the 1D electron gas with short-range interaction is algebraic, as expected. In this
section we calculate the associated exponent. 

We proceed in two stages. Initially we keep only the first cumulant $C_1$, and then afterwards we discuss the effect of $C_2$. In 1D it is possible to calculate the action (\ref{action}) exactly in the long-time asymptotic limit at the first-cumulant level. Setting $x_{\rm f}=x_{\rm i}=0$, we have [see (\ref{general result})] 
\begin{equation}
G(\tau_0) = {\rm const} \times G_0(\tau_0) \ \! e^{-S(\tau_{0})}.
\label{tree-level result}
\end{equation}
Considering a local interaction of the form $U(x-x') = U_0 \lambda \delta(x-x')$ the action is 
\begin{equation}
S(\tau_0) = {U_0 \lambda \over 2} \int_0^\beta \! d\tau \int_{-\infty}^\infty \! 
dx \ \! \big[ \rho(x,\tau) \big]^2.
\label{1D action}
\end{equation}
By linearizing the spectrum around the Fermi energy, the zero-temperature propagator at low energies is 
\begin{eqnarray}
G_0(x,\tau) &=& {1 \over \pi} \ \! {\rm Im} \ \! \bigg( {e^{i k_{\rm F}x} \over x 
+ i  v_{\rm F} \tau } \bigg) \nonumber \\
&=& {x \sin k_{\rm F}x - v_{\rm F} \tau \cos k_{\rm F}x \over \pi (x^2 + 
v_{\rm F}^2 \tau^2 )}.
\label{1D propagator}
\end{eqnarray}
The charge density (\ref{tunneling charge definition}) in this case is
\begin{eqnarray}
\rho(x,\tau) &=&  {v_{\rm F} \tau_0 \over \pi} \ \! {x \sin k_{\rm F}x - v_{\rm F} 
(\tau_0 - \tau) \cos k_{\rm F}x \over x^2 + v_{\rm F}^2 (\tau - \tau_0)^2} 
\nonumber  \\
&\times& {x \sin k_{\rm F}x - v_{\rm F} \tau \cos k_{\rm F}x \over x^2 + 
v_{\rm F}^2 \tau^2}.
\end{eqnarray}
Fig.\ (\ref{rhofig}) shows the charge density $\rho(x,\tau)$ as it spreads in time from its initially localized position. 

By a lengthy but straightforward calculation it can be shown that
\begin{equation}
S(\tau_0) = {3 \over 8 } {U_0 \lambda \over   v_{\rm F}\pi} \ln \bigg({\tau_0 
\over a}\bigg)  ,
\end{equation}
where $a$ is a microscopic cutoff. This leads to a power-law decay of 
the interacting propagator as
\begin{equation}
G(\tau) \sim {1 \over \tau^\alpha},
\end{equation}
where
\begin{equation}
\alpha = {3 \over 8}  {U_0 \lambda \over  v_{\rm F}\pi} + 1 
\end{equation}
is the propagator exponent. The DOS exponent $\delta$, defined as
\begin{equation}
N(\epsilon) = {\rm const} \times \epsilon^\delta,
\end{equation}
is given in this case by
\begin{equation}
\delta =  {3 \over 8 } {U_0 \lambda \over  v_{\rm F}\pi}.
\label{cumulant DOS exponent}
\end{equation}

The effects of the second cumulant $C_2$ are now straightforward to understand: In addition to introducing a slowly varying prefactor ${\cal A}$, whose only $\tau_0$ dependence comes from the time dependence of the screening in (\ref{Ueff definition}), the second cumulant screens the bare interaction and does not prevent the logarithmic divergence of the action, but it does modify the DOS exponent. It is interesting, however, that in the large $U$ limit, the effective interaction becomes independent of $U,$ a clear indication of nonperturbative behavior.

It is illustrative to compare (\ref{cumulant DOS exponent}) to the prediction of the perturbative x-ray edge limit of Ref.~\onlinecite{Patton&GellerPRB05}, where one neglects all field configurations in (\ref{exact form}) except $\phi_{\rm xr}$. There we found (for this same short-range interaction model),
\begin{equation}
\delta=2 \, {U_0 \lambda \over  v_{\rm F}\pi} 
\label{xray limit exponent}
\end{equation}
to leading order in $U.$  The DOS exponents (\ref{cumulant DOS exponent}) and (\ref{xray limit exponent}) are in qualitative agreement, but the inclusion of fluctuations about $\phi_{\rm xr}$ in (\ref{cumulant DOS exponent}) softens the exponent by almost a factor of four, as one might expect. 

The exact DOS exponent is not known for this model. In the next section we apply the cumulant method to the Tomonaga-Luttinger model, for which the exact propagator can be calculated using bosonization.

\subsection{Tomonaga-Luttinger model\label{TL model section}}
 
We consider the spinless or $U(1)$ Tomonaga-Luttinger model. The noninteracting spectrum is
\begin{equation}
\epsilon_k = \mu + v_{\rm F}(\pm k - k_{\rm F}),
\end{equation}
where the upper sign refers to the right branch and the lower to the left one.
The interaction is
\begin{eqnarray}
V &=& {1 \over 2} \int dx \ \delta n_i(x) \, U_{ij} \, \delta n_j(x) , \\ 
\delta n_{\pm}(x) &\equiv&  \lim_{a \rightarrow 0} : \! \psi_{\pm}(x+a) 
\psi_{\pm}(x) \! :,
\end{eqnarray} 
where the normal ordering is with respect to the noninteracting ground state.
The matrix $U$ has the form
\begin{equation}
{\bf U}=\left(\begin{array}{cc}U_{4} & U_{2} \\U_{2} & U_{4}\end{array}\right).
\end{equation}

We want to calculate
\begin{equation}
G_{\pm}(x_{\rm f} \tau_{\rm f}, x_{\rm i} \tau_{\rm i}) = - {\cal N} \, 
\langle T  \psi_{\pm}(x_{\rm f},\tau_{\rm f}) {\bar \psi}_{\pm}(x_{\rm i}, 
\tau_{\rm i}) \, e^{- \int d\tau V(\tau)} \rangle.
\end{equation}
Make a Hubbard-Stratonovich transformation of the form
\begin{equation}
e^{-{1 \over 2} \int \delta n_i \, U_{ij} \, \delta n_j} = {\int D\phi_{-} \,
D\phi_{+} \ e^{-{1 \over 2} \int \phi_i U_{ij}^{-1} \phi_j} \, e^{i\int \phi_i
\, \delta n_i} \over \int D\phi_{-} \, D\phi_{+} \ e^{-{1 \over 2} \int \phi_i
U_{ij}^{-1} \phi_j}},
\label{TL Hubbard-Stratonovich transformation}
\end{equation}
which leads to
\begin{align}
&G_{\pm}(x_{\rm f} \tau_{\rm f}, x_{\rm i} \tau_{\rm i}) =\nonumber \\ & {\cal N} \, 
{\int D\phi_{-} \, D\phi_{+} \ e^{-{1 \over 2} \int \phi_i U_{ij}^{-1} \phi_j}
\ g_{\pm}(x_{\rm f} \tau_{\rm f}, x_{\rm i} \tau_{\rm i} | \phi_{-},\phi_{+}) \over \int D\phi_{-} \, D\phi_{+} \ e^{-{1 \over 2} \int 
\phi_i U_{ij}^{-1} \phi_j}} ,
\label{TL exact form}
\end{align}
where
\begin{align}
&g_{\pm}(x_{\rm f} \tau_{\rm f}, x_{\rm i} \tau_{\rm i} |\phi_{+},\phi_{-} ) \equiv \nonumber \\  &-
\big\langle T \psi_{\pm}(x_{\rm f} \tau_{\rm f}) {\bar \psi}_{\pm}(x_{\rm i} 
\tau_{\rm i}) \, e^{i \int_0^\beta d \tau \int dx \, \phi_{i}(x,\tau) \, 
\delta n_{i}(x,\tau) } \big\rangle_0. 
\label{TL correlation function definition}
\end{align}
The correlation function (\ref{TL correlation function definition}) can also 
be written as 
\begin{equation}
g_{\pm}(x_{\rm f} \tau_{\rm f}, x_{\rm i} \tau_{\rm i}| \phi_{+},\phi_{-}) = g_{\pm}
(x_{\rm f} \tau_{\rm f}, x_{\rm i} \tau_{\rm i}| \phi_{\pm}) \cdot 
Z_{\mp}[\phi_{\mp}],
\end{equation}
where 
\begin{equation}
Z_{\pm}[\phi_{\pm}] \equiv \big\langle T e^{i \int_0^\beta d \tau \int dx \, \phi_{\pm}(x,
\tau) \, \delta n_{\pm}(x,\tau) } \big\rangle_{{0,\pm}},
\label{TL effective partition function}
\end{equation}
and
\begin{align}
&g_{\pm}(x_{\rm f} \tau_{\rm f}, x_{\rm i} \tau_{\rm i} |\phi_{\pm}) =\nonumber \\  &-
\big\langle T \psi_{\pm}(x_{\rm f} \tau_{\rm f}) {\bar \psi}_{\pm}(x_{\rm i} 
\tau_{\rm i}) \, e^{i \int_0^\beta d \tau \int dx \, \phi_{\pm}(x,\tau) \, 
\delta n_{\pm}(x,\tau) } \big\rangle_{{0,\pm}}. 
\label{reduced TL correlation function definition}
\end{align}
Next we cumulant expand both (\ref{TL effective partition function}) and (\ref{reduced TL correlation function definition}) to second order. For (\ref{TL effective partition function})
\begin{equation}
Z_{\pm}[\phi_{\pm}] \approx e^{\frac{1}{2}\int d \tau d\tau' \int dx dx' \,\Pi_{\pm}(x-x',\tau-\tau')\phi_{\pm}\phi'_{\pm}}
\end{equation}
Where $\Pi_{\pm}$ is the noninteracting density-density correlation function
\begin{equation}
\Pi_{\pm}(x,\tau) \equiv - \big\langle T \delta n_{\pm}(x,\tau) \, 
\delta n_{\pm}(0)  \big\rangle_{0,\pm} ,
\label{TL polarization definition}
\end{equation}
which can be written as
\begin{equation}
\Pi_{\pm}(x,\tau) = G_{0,\pm}(x, \tau) \, G_{0,\pm}(-x,-\tau).
\end{equation}
The noninteracting chiral propagator is
\begin{equation}
G_{0,\pm}(x, \tau) = \pm {1 \over 2 \pi i} \ {e^{\pm i k_{\rm F} x} \over x \pm
i v_{\rm F} \tau}.
\end{equation}
For (\ref{reduced TL correlation function definition})
\begin{align}
&g_{\pm}(x_{\rm f} \tau_{\rm f}, x_{\rm i} \tau_{\rm i} | \phi_{\pm})\approx G_{0,\pm}(x_{\rm f} \tau_{\rm f}, x_{\rm i} \tau_{\rm i} | \phi_{\pm}) \nonumber \\
& \times e^{\int dx d\tau\, C_{1,\pm}\phi_{\pm}+\int dxdx' d\tau d\tau'\, C_{2,\pm}\phi_{\pm}\phi'_{\pm}}
\end{align}
where
\begin{equation}
C_{1,\pm}(x,\tau)=-i{G_{0,\pm}(x_{\rm f}, x, \tau_0 - \tau) \, G_{0,\pm}(x, x_{\rm i}, \tau),
\over G_{0,\pm}(x_{\rm f}, x_{\rm i}, \tau_0)}
\end{equation}
and $C_{2,\pm}$ for this model reduces to
\begin{equation}
C_{2,\pm}(x-x',\tau-\tau')=\frac{1}{2}\,\Pi_{\pm}(x-x',\tau-\tau').
\end{equation}

Now we solve for $U_{\rm eff},$ defined by
\begin{align}
&\int dx''d\tau''{\bf U}_{\rm eff}^{-1}(x-x'',\tau-\tau''){\bf U}_{\rm eff}(x''-x',\tau''-\tau')=\nonumber \\
&\delta(x-x')\delta(\tau-\tau')\openone,
\label{TL Ueff}
\end{align}
where
\begin{align}
&U^{\rm eff}_{ij}(x,\tau)=\nonumber \\ &\left[{U}_{ij}^{-1}(x,\tau)-{\Pi}_{\pm}(x,\tau)\delta_{i+}\delta_{j+}-{\Pi}_{\mp}(x,\tau)\delta_{i-}\delta_{j-}\right]^{-1}
\end{align}
and $\openone$ is a $2\times 2$ identity matrix. To achieve this we Fourier transform (\ref{TL Ueff}). This reduces (\ref{TL Ueff}) to a matrix equation which gives
\begin{equation}
{\bf U}_{\rm eff}(k,\omega)=\left[ \left(\begin{array}{cc}U_{4} & U_{2} \\U_{2} & U_{4}\end{array}\right)^{-1}-\left(\begin{array}{cc}\Pi_{\pm}(k,\omega) & 0 \\0 & \Pi_{\mp}(k,\omega)\end{array}\right)\right]^{-1}.
\end{equation}
The $++$ or $--$ component is 
\begin{equation}
U_{\rm eff}(k,\omega)=\frac{U_{4}-(U^{2}_{4}-U^{2}_{2})\Pi_{\mp}(k,\omega)}{1-U_{4}\Pi_{t}(k,\omega)+(U^{2}_{4}-U^{2}_{2})\Pi_{+}(k,\omega)\Pi_{-}(k,\omega)}
\end{equation}
where
\begin{equation}
\Pi_{\rm t}(k,\omega) =\Pi_{+}(k,\omega)+\Pi_{-}(k,\omega)=
- {1 \over \pi} \, { k^2 \over (\omega + ik) (\omega - ik) }.
\end{equation}
The effective interaction for right movers is (with $v_{\rm F}=1$)
\begin{equation}
U_{\rm eff}(k,\omega) = U_4 \, {(\omega + ik) ( \omega - i u k)  \over 
(\omega + i v k) (\omega - i v k) },
\end{equation}
where
\begin{equation}
u \equiv 1 + {(U_4 / 2 \pi)^2 - (U_2/2 \pi)^2 \over (U_4 / 2 \pi)} 
\end{equation}
and
\begin{equation}
v \equiv \sqrt{ \bigg( 1 + {U_4 \over 2 \pi} \bigg)^2 - 
\bigg( {U_2 \over 2 \pi} \bigg)^2 }.
\end{equation}
The action can be written as
\begin{equation}
S = {1 \over 2} \int {dk \over 2 \pi} {d \omega \over 2 \pi} \ 
\rho_{\pm}(-k, -\omega) \, U_{\rm eff}(k,\omega) \, \rho_{\pm}(k, \omega).
\label{TL action}
\end{equation}
The chiral tunneling charge density is
\begin{equation}
\rho_{\pm}(x,\tau) = - {G_{0,\pm}(x_{\rm f}, x, \tau_0 - \tau) \, G_{0,\pm}(x, x_{\rm i}, \tau)
\over G_{0,\pm}(x_{\rm f}, x_{\rm i}, \tau_0)}.
\end{equation}

We now specialize to the DOS case where $x_{\rm i} = x_{\rm f} = 0$ and assuming  
right movers we set $\rho_{+}\equiv \rho$. The tunneling charge density is
\begin{equation}
\rho(x, \tau ) = {v_{\rm F} \tau_0 \over 2 \pi}{1 \over (x + i v_{\rm F} \tau)
[x + i v_{\rm F} (\tau - \tau_0)]}, 
\end{equation}
which satisfies
\begin{equation}
\int dx \ \rho(x, \tau ) = \Theta(\tau) \, \Theta(\tau_0 - \tau).
\end{equation} 

\begin{widetext}

Fourier transforming, we find that
\begin{equation}
\rho(k,\omega) = {1 \over i \omega - v_{\rm F}k} \big[ \big(e^{i \omega 
\tau_0} -1 \big) \, \Theta(k) + \big( e^{i \omega \tau_0} - e^{v_{\rm F} k 
\tau_0} \big) \Theta(-k) \big]
\end{equation}
and
\begin{equation}
\rho(k,\omega) \, \rho(-k,-\omega) = {1 \over (\omega +ik)^2} \bigg[ \bigg(
1 - e^{i \omega \tau_0} e^{- k \tau_0} + e^{-k \tau_0} - e^{-i \omega \tau_0}
\bigg) \Theta(k) + \bigg(1 - e^{i \omega \tau_0} + e^{k \tau_0} - 
e^{-i \omega \tau_0}  e^{k \tau_0} \bigg) \Theta(-k) \bigg].
\end{equation}
The action therefore is
\begin{align}
S(\tau_{0})=&{1 \over 2} \int {dk \over 2 \pi} {d \omega \over 2 \pi}\, {(\omega + ik) ( \omega - i u k)  \over 
(\omega + i v k) (\omega - i v k) } \ {U_4  \over (\omega +ik)^2}\nonumber \\ &\times \bigg[ \bigg(
1 - e^{i \omega \tau_0} e^{- k \tau_0} + e^{-k \tau_0} - e^{-i \omega \tau_0}
\bigg) \Theta(k) + \bigg(1 - e^{i \omega \tau_0} + e^{k \tau_0} - 
e^{-i \omega \tau_0}  e^{k \tau_0} \bigg) \Theta(-k) \bigg].
\end{align}
The action maybe written as $S=S_{>}+S_{<}$ where
\begin{equation}
\label{pos action}
S_{>}=\frac{U_{4}}{8 \pi^2}\int\limits_{0^+}^{\infty}dk\,\int\limits_{-\infty}^{\infty}d\omega \, { ( \omega - i u k)  \over 
(\omega + i v k) (\omega - i v k) (\omega+ik)}\bigg(
1 - e^{i \omega \tau_0} e^{- k \tau_0} + e^{-k \tau_0} - e^{-i \omega \tau_0}
\bigg)
\end{equation}
and
\begin{equation}
S_{<}=\frac{U_{4}}{8 \pi^2}\int\limits_{-\infty}^{0^-}dk\,\int\limits_{-\infty}^{\infty}d\omega \,\, { ( \omega - i u k)  \over 
(\omega + i v k) (\omega - i v k) (\omega+ik)}\bigg(1 - e^{i \omega \tau_0} + e^{k \tau_0} - 
e^{-i \omega \tau_0}  e^{k \tau_0} \bigg).
\end{equation}
\end{widetext}
$S_{>}=S_{<}$ under change of coordinates $k\rightarrow -k$ and $\omega \rightarrow -\omega$, so $S=2S_{>}$.
To proceed we need the large-$\tau_0$ asymptotic result
\begin{equation}
I(\tau_0) \equiv \int_{0^+}^\infty dk \ {e^{-k \tau_0} \over k}
\longrightarrow - \ln \tau_0
\end{equation}
where the additive constant, not shown explicitly, is cutoff dependent.  These lead, in the large $\tau_0$ limit, to
\begin{equation}
S=\frac{U_4}{4\pi}\left[\frac{2(1+u)}{(1+v)(1-v)}-\frac{u+v}{v(1-v)}\right]\ln(\tau_{0})
\end{equation}
or 
\begin{equation}
S=\frac{U_4}{4\pi}\left[\frac{v-u}{v(1+v)}\right]\ln(\tau_{0}).
\end{equation}
Finally, we obtain
\begin{equation}
S = \delta \ln \tau_0 +  {\rm const} + O(1/\tau_0)
\end{equation}
and
\begin{equation}
N(\epsilon) = {\rm const} \times \epsilon^\delta ,
\end{equation}
where
\begin{eqnarray}
\delta &=&  {U_4 \over 4 \pi} \ {v-u \over v(1+v)} = {(u-v)(1-v) 
\over 2 v(1+u)} \\ 
&=& {1 + {U_4 \over 2 \pi v_{\rm F}} - \sqrt{ \big(1 + {U_4 \over 2 \pi 
v_{\rm F}} \big)^2 - \big({U_2 \over 2 \pi v_{\rm F}} \big)^2 } \over 2 \sqrt{ 
\big(1 + {U_4 \over 2 \pi v_{\rm F}} \big)^2 - \big({U_2 \over 2 \pi 
v_{\rm F}} \big)^2 }} .
\end{eqnarray}
This is in exact agreement with the bosonization result
\begin{equation}
\delta = {g + g^{-1} \over 2} - 1
\end{equation}
with
\begin{equation}
g = \sqrt{{1 + {U_4 \over 2 \pi v_{\rm F}} -  {U_2 \over 2 \pi v_{\rm F}} \over
1 + {U_4 \over 2 \pi v_{\rm F}} + {U_2 \over 2 \pi v_{\rm F}}}}.
\end{equation}

Why does the cumulant method give the exact result for this model? The answer is that a second order cumulant expansion of the form used here is exact for free bosons, which are the exact eigenstates of the Tomonaga-Luttinger model.\cite{vignalenote}

\section{DISCUSSION}

Our principal result (\ref{general result}) suggests that the dominant effect of interaction on the low-energy DOS in a variety of strongly correlated electron systems is to add a semiclassical time-dependent charging energy contribution to the total potential barrier seen by a tunneling electron, as in Ref.~\onlinecite{Levitov97}. The energy is computed according to classical electrostatics with a dynamically screened two-particle interaction. In 2D and 3D the added charge is accommodated efficiently and reaches a zero-action state at long times. In 1D the added charge leads to diverging action, and hence suppressed tunneling.

The robustness of our method has not been fully explored, although it is known to fail qualitatively in systems with ground state degeneracy, such as in the quantum Hall fluid. We believe the cause of this failure to be the non-satisfiability of the sum rule (\ref{sum rule}) in such situations. In the future we plan to apply this method to other exactly solvable systems, such as the 1D Hubbard and Calogero-Sutherland models.

\acknowledgments

This work was supported by the National Science Foundation under CAREER Grant No.~DMR-0093217, and by a Cottrell Scholars Award from the Research Corporation. It is a pleasure to thank Phil Anderson, Matthew Grayson, Dmitri Khveshchenko,  Allan MacDonald, Emily Pritchett, David Thouless, Shan-Ho Tsai, and Giovanni Vignale for useful discussions. M.G. would also like to acknowledge the Aspen Center for Physics, where some of this work was carried out. 

\bibliography{/Users/mgeller/Papers/bibliographies/MRGhall,/Users/mgeller/Papers/bibliographies/MRGmanybody,/Users/mgeller/Papers/bibliographies/MRGbooks,/Users/mgeller/Papers/bibliographies/MRGgroup,cumulantnotes}

\begin{thebibliography}{21}
\bibitem[{ste()}]{stepfunctionnote}
\bibinfo{note}{$\Theta$ is the Heaviside step function.}
\bibitem[{fun()}]{functionalnote}
\bibinfo{note}{Note that $S$ is not a functional of $\rho$, but rather is a
  function of $\tau_0$.}
\bibitem[{vig()}]{vignalenote}
\bibinfo{note}{This interesting observation was pointed out to us by Giovanni
  Vignale.}
\end{thebibliography}

\end{document}